\documentclass[aps,pra,superscriptaddress,amsmath,amssymb,preprintnumbers,showpacs]{revtex4}
\usepackage{amssymb}
\usepackage{graphicx}
\usepackage{bm}
\usepackage{empheq}
\usepackage{color}
\usepackage{upgreek}
\usepackage{cancel}
\usepackage{slashed}
\usepackage{soul}

\newcommand{\Ignore}[1]{}
\newcommand{\ToDelete}[1]{{\color{red} #1}}

\renewcommand{\ToDelete}[1]{}

\newcommand{\Ket}[1]{\left\vert #1\right\rangle}
\newcommand{\Bra}[1]{\left\langle #1\right\vert}

\newcommand{\BraKet}[2]{\left\langle#1\vert #2\right\rangle}

\newcommand{\FO}{F_{mn;m'n'}^{(1)} (t)}
\newcommand{\FTW}{F_{mn;m'n'}^{(2)}(t)}
\newcommand{\FTH}{F_{mn;m'n'}^{(3)}(t)}
\newcommand{\FF}{F_{mn;m'n'}^{(4)}(t)}
\newcommand{\FFI}{F_{mn;m'n'}^{(5)}(t)}
\newcommand{\WO}{W_{mn;m'n'}^{(1)}}
\newcommand{\WTW}{W_{mn;m'n'}^{(2)}}
\newcommand{\WTH}{W_{mn;m'n'}^{(3)}}
\newcommand{\WF}{W_{mn;m'n'}^{(4)}}
\newcommand{\nnp}{n'}
\newcommand{\mmp}{m'}
\newcommand{\Bess}{J_{m}(k_{mn}r)}
\newcommand{\Bessp}{J_{m'}(k_{m'n'}r)}
\renewcommand{\eqref}[1]{Eq.~(\ref{#1})}

\newcommand{\vct}[1]{\mathbf{#1}}
\newcommand{\dom}{{\cal D}}
\newcommand{\border}[1]{\partial {#1}}
\newcommand{\compl}[1]{\bar{#1}}

\def\ii{\mathrm{i}}

\begin{document}

\title{Dynamics of a particle confined in a two-dimensional dilating and deforming domain}

\author{Fabio Anz\`{a}}
\address{Dipartimento di Fisica e Chimica, Universit\`a di Palermo, I-90123 Palermo, Italy}
\address{Department of Atomic \& Laser Physics, Clarendon Laboratory, University of Oxford, Parks Road, Oxford OX1 3PU, UK}

\author{Sara Di Martino}
\address{Dipartimento di Matematica, Universit\`a di Bari, I-70125  Bari, Italy}
\address{School of Mathematical Sciences, The University of Nottingham, University Park, Nottingham NG7 2RD, United Kingdom}

\author{Antonino Messina}
\address{Dipartimento di Fisica e Chimica, Universit\`a di Palermo, I-90123 Palermo, Italy}

\author{Benedetto Militello}
\address{Dipartimento di Fisica e Chimica, Universit\`a di Palermo, I-90123 Palermo, Italy}

\begin{abstract}

Some recent results concerning a particle confined in a
one-dimensional box with moving walls are briefly reviewed. By
exploiting the same techniques used for the 1D problem, we
investigate the behavior of a quantum particle confined in a
two-dimensional box (a 2D billiard) whose walls are moving, by recasting the
relevant mathematical problem with moving boundaries in the form
of a problem with fixed boundaries and time-dependent Hamiltonian.
Changes of the shape of the box are shown to be important, as it
clearly emerges from the comparison between the \lq\lq
pantographic\rq\rq\, case (same shape of the box through all the
process) and the case with deformation.

\end{abstract}

%\pacs{03.65.-w, 03.65.Db, 03.65.Xp}


\maketitle

\section{Introduction}\label{sec:Introduction}

Over the decades several authors have dealt with the problem of
physical systems with moving boundaries, both in quantum field
theory, especially in connection with the Casimir
effect~\cite{ref:CasimirReview,ref:Wilson2010}, and in the context
of quantum mechanics, for example in connection with the Fermi
problem of a quantum bouncer~\cite{ref:Fermi}. Over the years,
several works appeared studying the problem of particles confined
in a box with moving walls~\cite{ref:Doescher, ref:Pinder,
ref:Schlitt, ref:Dodonov, ref:DiMartino}, sometimes focusing
on boundaries having specific shapes~\cite{ref:Lenz2011,
ref:Mousavi2012, ref:Mousavi2013}. The study of this kind of
problems is relevant to several conceptual aspects of quantum
mechanics, from the analysis of the semiclassical limit of a
quantum (chaotic)
system~\cite{ref:BinKang1992,ref:BinKang1993,ref:BinKang1994,ref:Schulman1994}
to the incoming of geometric phases~\cite{ref:Zou2000}, and it is
connected with the derivation of analytical solutions of the
dynamics of systems governed by such mathematically complicated
potentials as delta functions~\cite{ref:Gaveau1986}. The interest
in such a class of problems is present in different physical
scenarios and can lead to intriguing applications.
For example, in the field of cavity quantum electrodynamics appeared a study of the implications
of small displacements of the mirrors on the dynamics of an atom
interacting with the cavity modes~\cite{ref:Garraway2008}. On the
other hand, cooling techniques of trapped particles based upon
exploitation of expanding boxes have been
proposed~\cite{ref:XiChen2009}. Furthermore,
the study of two particles in a box with moving boundaries has been
recently presented, and the relevant results suggest the
idea of an effective interaction between the particles induced by their
common interaction with the moving walls~\cite{ref:Mousavi2014}. 
The problem of many particles in a box with moving walls 
(possibly in higher dimensional situations) has also been treated, 
especially in connection with shortcuts to adiabaticity and in the 
context of Bose-Einstein condensates~\cite{ref:delCampo2012,ref:delCampo2013}.

In this paper, we firstly review some recent results concerning
the dynamics of a particle confined in a one-dimensional box,
pointing out the delicate mathematical aspects of this class of
problems and presenting an approach to overcome the relevant
difficulties~\cite{ref:DiMartino}. Secondly, on the basis of this 
approach, we report on the analytical
study of a wide class of problems of a single particle in a
two-dimensional box whose walls are moving. In fact, following the
way paved in Ref.~\cite{ref:Lenz2011} for a 2D elliptical billiard
and in Ref.~\cite{ref:DiMartino} for a 1D box, we transform the
original problem of a system with changing boundaries into the
problem of a system with fixed boundaries governed by a
time-dependent Hamiltonian. Generally speaking, resolution of
dynamical problems in the presence of time-dependent Hamiltonians
is a hard task, hence complete or partial analytical solutions
are known only in special cases,  for example in
the adiabatic limit~\cite{ref:Messiah} or in the presence of periodical
potentials~\cite{ref:Traversa2013,ref:Moskalets2002,ref:Shirley1965}.
In most cases, perturbation
theory~\cite{ref:Aniello2005,ref:Militello2007,ref:Rigolin2008,ref:Zagury2010}
or numerical
methods~\cite{ref:Shao2009,ref:VanDijk2011,ref:Varga2012} are the
only effective ways to solve such problems.

In our case, we will exploit standard perturbation approach in
order to bring to the light some new effects. We will show that
the very novelty in the study of 2D and 3D problems with respect
to the one-dimensional case is the role of the shape of the moving
box, which is obviously absent at all in the one-dimensional case.
In fact, for two- and three-dimensional boxes, we can consider two
classes of problems: on the one hand the ones that we dub
as \lq\lq pantographic\rq\rq,
where the dimensions of the box change but its shape remains invariant; on the
other hand the cases where the shape (and possibly the dimensions) of the box
changes, meaning that we are in the presence of \lq\lq
deformation\rq\rq of the contour. We start our analysis by
constructing the framework of the most general case of a 2D box
undergoing dilation and deformation, then we proceed by
specializing our analysis to the pantographic case and eventually
we present a perturbative treatment of the problem where the shape of
the boundary is only slightly modified (small deformations).
Though in Refs.~\cite{ref:Mousavi2012} and \cite{ref:Mousavi2013}
three-dimensional dilating domains have been considered, the role
of deformation has not been treated in such works. On the
contrary, in Ref.~\cite{ref:Lenz2011} it is analyzed an elliptical
billiard whose axes have time-dependent lengths, focusing in
particular on the case where the lengths of the axes are
oscillating at a given frequency, and then numerically
evaluating the dynamics of the
particle in the moving box. Albeit the
authors succeed in going through many details, their analysis is
strictly related to a specific geometry.
Our work is different since we aim at describing the very more general situation wherein the shape can be
modified in any possible way, provided the domain is kept as a
star domain. In fact, we will write down the relevant general
equations and will study them under the assumption that the
deformation can be considered as a perturbation. On this basis, we
will evaluate the time evolution through the standard approach in
a specific case (a circle which is deformed to an ellipse).

The paper is organized as follows. In the next section we
summarize some recent results related to the one-dimensional
problem, also presenting the general approach of
Ref~\cite{ref:DiMartino} (which is analogous to that of
Ref.~\cite{ref:Lenz2011}) that we will use through all the paper.
In section \ref{sec:2DHamiltonian} we introduce the problem of a
two-dimensional box whose border is changing, defining the two
classes of pantographic changes and changes with deformation. In
section \ref{sec:Pantography} we study the pantographic case,
while in section \ref{sec:Deformations} we use perturbation theory to study the effects of deformation of the border of
the box. Finally in section \ref{sec:Discussion} we provide some
conclusive remarks and very briefly discuss the extension of the
results to the three-dimensional case.

\section{Summary of previous results: The one-dimensional problem}

\subsection{General framework}

A (free) non-relativistic quantum particle of mass $\mu>0$
confined in a certain domain $\dom$ is formally described by the
usual (free) particle Hamiltonian in the space region
corresponding to the box,
\begin{equation}
  H = -\frac{\hbar^2}{2 \mu} \nabla^2\,,
\end{equation}
imposing suitable boundary conditions, in such a way that $H$ is self-adjoint. One of the possible is that the wave function
of the particle is vanishing in the border of the box
($\border{{\dom}}$):
\begin{equation}\label{eq:BoundaryCondition}
\psi(\vct{r})=0\,,\,\,\,\,\, \forall\, \vct{r} \in
\border{{\dom}}\,.
\end{equation}

Now, if the domain $\dom$ is not static, there is a non trivial
technical problem in the resolution of the Schr\"odinger equation
related to the fact that we have to solve a differential equation
in a Hilbert space which is continuously changing, so that, for
example, the time derivative of the wave function is not well
defined in the border of the box. Indeed, if $\vct{r}_0$ is a
point of the border at time $t+\mathrm{d}t$ which does not belong
to the border at time $t$, then the quantity $(\psi(\vct{r}_0,
t+\mathrm{d}t) - \psi(\vct{r}_0, t))/\mathrm{d}t$ is meaningless.
According to the approach in Ref~\cite{ref:DiMartino}, this
difficulty can be overcome by enlarging the domain of definition
of the Hamiltonian in such a way that the such operator acts as
the usual (free) particle Hamiltonian in the space region
corresponding to the box and is zero elsewhere (i.e., in the
complement $\compl{\dom}$):
\begin{equation}\label{eq:ExtendedHamiltonian}
  H_{\cal D} = -\frac{\hbar^2}{2 \mu} \nabla^2\oplus_{\compl{\dom}}
  0\,.
\end{equation}
Of course, the condition expressed by \eqref{eq:BoundaryCondition}
that the wave function of the particle vanishes in the border of
the box will be kept. In this way, the wave function turns out to
be properly defined for any $\vct{r}$, being zero out of the box.
Though the most natural choice for the extension of the
Hamiltonian would be to put the operator equal to infinity out of
the box  (which is also a better description from the physical
point of view), we stress that here the main reason for extending
the Hamiltonian operator is that we want to extend the wave
function out of the box. Now, since the wave function is anyway
vanishing out of the box because of the boundary conditions, any
extension of the Hamiltonian would be fine, in the sense that the
specific choice of the form of the operator out of the box is
irrelevant to the dynamical description of the system. Therefore,
we have decided to exploit the simplest mathematical extension.

In order to better treat the dynamical problem associated to
\eqref{eq:ExtendedHamiltonian}, we can map it into the problem of
a particle in a fixed domain, say $\dom_0$, governed by a suitable
effective time-dependent Hamiltonian, $H_{\mathrm{eff}}$:
\begin{equation}
%  H_{\dom_0} = -\frac{\hbar^2}{2 \mu} \nabla^2\oplus_{\compl{\dom_0}} 0\,,
  H_{\dom_0} = H_{\mathrm{eff}}\oplus_{\compl{\dom}_0} 0\,,
\end{equation}
\begin{equation}\label{eq:BoundaryConditionStatic}
\phi(\vct{r})=0\,,\,\,\,\,\, \forall\, \vct{r} \in
\border{\dom_0}\, .
\end{equation}

This passage can be done through the action of a suitable unitary
operator.

\subsection{The one-dimensional box}

In Ref~\cite{ref:DiMartino} it has been studied the case of a
particle in a one-dimensional box when both of its walls are
moving, each one with its own velocity; stated another way, the
size of the box is changing and at the same time the center of the
box is moving. For the sake of simplicity, we will summarize the
results when the center of the box is quiet, so that the particle
domain is $\dom=[-R x_0/2,R x_0/2]$, with $R$ the dilation
dimensionless parameter. This problem is then mapped into the one
related to the box delimited by $\dom_0=[-x_0/2,x_0/2]$. The
relevant transformation is given by the unitary operator $U$ such
that
\begin{equation}
  \phi(x) = (U \psi)(x) = R^{1/2} \psi(R x)\,.
\end{equation}
Under this transformation, the original Hamiltonian,
\begin{equation}
  H_{\dom} = -\frac{\hbar^2}{2 \mu} \frac{\partial^2}{\partial x^2} \oplus_{\compl{\dom}} 0 \,,
\end{equation}
is replaced by the new generator of the time evolution:
\begin{equation}
  H_{\dom_0} = \left[ -\frac{\hbar^2}{2 \mu R^2} \frac{\partial^2}{\partial x^2} +
  \ii\hbar \frac{\dot{R}}{R}\left(\frac{1}{2}+x\frac{\partial}{\partial_x}\right)\right] \oplus_{\compl{\dom}_0} 0\,.
\end{equation}

The price to pay to obtain static boundaries is to make the
Hamiltonian explicitly time-dependent. In particular, in this case we have
obtained that the kinetic energy, in the new picture, is that of a
particle with changing mass (as an effect of the presence of the
dilation parameter $R$ in the denominator). Moreover, the particle
is subjected to an additional energy term, the one proportional to
$\dot{R}/R$, which is nothing but $\ii \dot{U}U^\dag$, i.e. the
generator of the dilation. It is worth noting that among
the two terms, only the first one preserves the physical meaning of energy
of the particle, being the result of a unitary transformation of
the original energy operator. The second term, instead, is
relevant only for determining the dynamics of the particle in the
new picture.

Now, since the term $x\partial_x$ does not commute with the
operator $\partial_x^2$, the kinetic energy of the particle is not
conserved during the process. The relevant energy rate equation,
which generalizes and improves the result present in
Ref~\cite{ref:Konishi}, is:
\begin{equation}
\dot{E}(t) = - \frac{\hbar^2 \dot{R}}{2\mu R^3} \left[ |
\phi'(1/2, t)|^2 + |\phi'(-1/2, t)|^2\right] \,.
\end{equation}

This interesting formula shows that the change of energy is
determined by the contact between the particle and the walls. One
could think that a natural generalization of this would still be
valid in a two- or three-dimensional context. This is what will be
shown in the next section, together with some specific properties
which have no analogous in the 1D case.

\section{A quantum particle in a two-dimensional box}\label{sec:2DHamiltonian}

In order to start our analysis of a two-dimensional problem, let
us consider the case in which the domain of the wave functions is
a star domain lying on a plane and delimited by a curve $\gamma$.
We recall here that a region $S$ of $\mathbb{R}^n$ is said to be a
star domain if there exists a point $x_0\in S$ such that $\forall
x\in S$ the segment $\lambda x_0 +(1-\lambda) x$, for $\lambda\in
[0,1]$, lies in the interior of $S$. Assume that the curve
describing the walls of the box can be represented by the
following equation:
\begin{equation}
  r = \gamma(\theta)\,,\qquad \theta \in [0, 2\pi]\,,
\end{equation}
where the origin of the polar coordinates is the {\it center} (or
one of the possible centers) of the domain. Here the one-to-one
correspondence between $r$ and $\theta$ is guaranteed by the
property of the domain to have a star-shape. The region delimited
by $\gamma$ is expressible as $\dom = S_\gamma = \{P(r,\theta)
\,\, | \,\, r \le \gamma(\theta), \theta \in [0, 2\pi] \}$. The
boundary conditions on the wave function $\psi(r, \theta)$ are
expressible as:
\begin{equation}
  \psi(\gamma(\theta), \theta) = 0\,, \qquad \forall\,\theta\in [0, 2\pi].
\end{equation}

The domain $S_\gamma$ can be mapped to another domain ($\dom_0 =
S_\eta = \{P(r,\theta) \,\, | \,\, r \le \eta(\theta)\,, \theta
\in [0, 2\pi]$) by the transformation $U$ acting as follows:
\begin{equation}
  \phi(s, \theta) = (U \psi)(s, \theta) = R(\theta) \psi(s R(\theta), \theta)
  \,,\qquad R(\theta) = \gamma(\theta)/\eta(\theta)\,.
\end{equation}
Such transformation is unitary. Indeed,
\begin{eqnarray}
  \nonumber
  \int_{S_\eta} \phi_1^*(s, \theta)\phi_2(s, \theta) \, s \, \mathrm{d}s \, \mathrm{d}\theta &=& %
  \int_{S_\eta} R(\theta)^2 %
         \psi_1^*(s R(\theta), \theta)\psi_2(s R(\theta), \theta) \, s \, \mathrm{d}s \, \mathrm{d}\theta \\
  &=& %
  \int_{S_\gamma} %
         \psi_1^*(r, \theta)\psi_2(r, \theta) \, r \, \mathrm{d}r \, \mathrm{d}\theta \, .
\end{eqnarray}

If the boundary is moving, we have that the curve $\gamma$ changes
and then the function $R$ depends on $t$, $R(\theta, t)$.

Now, for the 2D problem of a confined particle, we have:
%\begin{subequations}
\begin{eqnarray}
  H &=& \frac{p^2}{2\mu} = -\frac{\hbar^2}{2\mu} \nabla^2\,,\\
  \nabla^2 &=& \left( %
  \frac{1}{r}\frac{\partial}{\partial r} +
  \frac{\partial^2}{\partial r^2} +
  \frac{1}{r^2}\frac{\partial^2}{\partial\theta^2}
  \right)\,,
\end{eqnarray}
%\end{subequations}

\begin{equation}
  (U^\dag \phi)(r,\theta)= \frac{1}{R(\theta, t)} \, \phi(r/R(\theta, t), \theta)\,,
\end{equation}

\begin{eqnarray}
  \nonumber
  \left(\frac{\mathrm{d}U}{\mathrm{d}t} \psi\right)(r,\theta) &=&
  \frac{\mathrm{d}}{\mathrm{d}t} (U \psi)(r,\theta) = \\
  \nonumber
  &=& \frac{\mathrm{d}}{\mathrm{d}t} \left[R(\theta,t) \psi(r
  R(\theta,t),\theta)\right] = \dot{R}(\theta, t) \psi(r
  R(\theta,t),\theta) + R(\theta, t) r \dot{R}(\theta, t) \psi_1(r
  R(\theta,t),\theta) \\
  &=& \dot{R}(\theta, t) \left[ \psi(r
  R(\theta,t),\theta) + r R(\theta, t) \psi_1(r
  R(\theta,t),\theta)\right]
  \,,
\end{eqnarray}
where $\psi_k$ is the derivative with respect to the $k$-th
argument. Then,
\begin{eqnarray}
  \nonumber
  \ii\hbar \left(\frac{\mathrm{d}U}{\mathrm{d}t} \left( U^\dag \phi \right)\right)(r,\theta)
  &=&
  \ii\hbar \frac{\mathrm{d}U}{\mathrm{d}t} \left[ \frac{1}{R(\theta, t)}\phi(r/R(\theta,
  t), \theta) \right]\\
  \nonumber
  &=& \ii\hbar \frac{\dot{R}(\theta, t)}{R(\theta, t)} \left[ \phi(r,\theta) + r
  \phi_1(r,\theta)\right]\\
  &=& \ii\hbar \frac{\dot{R}(\theta, t)}{R(\theta, t)} \left( 1 + \, r \frac{\partial}{\partial r}
  \right) \phi(r,\theta)
  \,,
\end{eqnarray}

\begin{eqnarray}
  \nonumber
  U H U^\dag \phi(r, \theta) &=& %
  -\frac{\hbar^2}{2\mu} \, U \left(  \frac{1}{r}\frac{\partial}{\partial r} +
  \frac{\partial^2}{\partial r^2} + \frac{1}{r^2}\frac{\partial^2}{\partial\theta^2} \right) \frac{1}{R} \phi\left(r/R, \theta\right) \\
  \nonumber
  &=& %
  -\frac{\hbar^2}{2\mu} \, \Bigg\{
  \frac{1}{r R^2} \frac{\partial}{\partial r} \phi(r, \theta) + %
  \frac{1}{R^2} \frac{\partial^2}{\partial r^2} \phi(r, \theta) +
  \frac{1}{r^2 R^2} \frac{\partial^2}{\partial \theta^2}\phi(r, \theta) \\ %
  \nonumber
  &+& %
  \frac{1}{R} \left(\frac{1}{R}\right)_{\theta\theta} \frac{1}{r^2} \phi(r, \theta) + %
  \left[ \left(\left(\frac{1}{R}\right)_{\theta}\right)^2 +
  \left(\frac{1}{R}\left(\frac{1}{R}\right)_{\theta}\right)_\theta \right] \frac{1}{r} \frac{\partial}{\partial r}
  \phi(r, \theta) %
  \\ %
  &+& %
  \frac{2}{R} \left(\frac{1}{R}\right)_{\theta} \frac{1}{r^2} \frac{\partial}{\partial \theta} \phi(r, \theta) + %
  \left[\left(\frac{1}{R}\right)_{\theta}\right]^2 \frac{\partial^2}{\partial r^2} \phi(r, \theta) + %
  \frac{2}{R} \left(\frac{1}{R}\right)_{\theta} \frac{1}{r} \frac{\partial}{\partial r} \frac{\partial}{\partial \theta} \phi(r, \theta)
  \Bigg\}
  \,,\label{eq:Hamilt_Transformed_Deformation}
\end{eqnarray}
where the subindex $\theta$ means derivative with respect to
$\theta$. It is worth mentioning two technical aspects relevant to
the previous equations. First, note that the wave functions $\psi$
and $\phi$ are assumed as time-independent. Indeed, they are not
meant as solutions of the Schr\"odinger equation, but as generic
elements of the Hilbert spaces to which they belong.
We just use them to make explicit the action of the operators considered above. Second, when we perform
$\theta$-derivatives, we have to take care of the twofold
dependence on $\theta$ of the wave function, due to both the \lq
natural\rq\, $\theta$-dependence and the possible additional
$\theta$-dependence due to the action of $U$ or $U^\dag$. For
example, $\partial_\theta U^\dag \phi(r,\theta) = \partial_\theta
[R^{-1}\phi(R^{-1}r,\theta)]$, where $R$ in general is a function
of $\theta$.

The effective Hamiltonian associated to the problem in the fixed
domain \, --- \, $H_{\mathrm{eff}} = U H U^\dag + \ii\hbar
\dot{U}U^\dag$ \, --- \, is rather involved, due to the
deformation of the boundary. In fact, such Hamiltonian can be
considered as the sum of three terms:
\begin{eqnarray}
  H_{\mathrm{eff}} &=&  %U H U^\dag + \ii\hbar\dot{U} U^\dag =
    H_\mathrm{1} + H_\mathrm{2}+ H_\mathrm{3}\,,
  \label{eq:intro_Heff}
  \\
  H_\mathrm{1} &=&   -\frac{\hbar^2}{2\mu R^2} \nabla^2 \,,
  \label{eq:intro_Hvm}
  \\
  H_\mathrm{2} &=&   \ii\hbar \frac{\dot{R}}{R} \left(1 + \, r \frac{\partial}{\partial
  r}\right)\,,
  \label{eq:intro_Hdil}
  \\
  \nonumber
  H_\mathrm{3} &=&   %
  -\frac{\hbar^2}{2\mu}\Bigg\{\frac{1}{R} \left(\frac{1}{R}\right)_{\theta\theta} \frac{1}{r^2} + %
  \left[ \left(\left(\frac{1}{R}\right)_{\theta}\right)^2 +
  \left(\frac{1}{R}\left(\frac{1}{R}\right)_{\theta}\right)_\theta \right] \frac{1}{r} \frac{\partial}{\partial r}
  \\ %
  &+& %
  \frac{2}{R} \left(\frac{1}{R}\right)_{\theta} \frac{1}{r^2} \frac{\partial}{\partial \theta}  + %
  \left[\left(\frac{1}{R}\right)_{\theta}\right]^2 \frac{\partial^2}{\partial r^2}  + %
  \frac{2}{R} \left(\frac{1}{R}\right)_{\theta} \frac{1}{r} \frac{\partial}{\partial r} \frac{\partial}{\partial \theta}
  \Bigg\}
  \,.
  \label{eq:intro_Hdef}
\end{eqnarray}

It is worth noting that when the relation between the moving
domain $\dom$ and the static domain $\dom_0$ is pantographic,
which means that they have the same shape, the function
$R(\theta,t)$ does not really depend on $\theta$ and all the terms
in $H_\mathrm{3}$ vanish. On the contrary, if there is a
deformation of the domain, $R_\theta$ is non zero and
$H_\mathrm{3}$ is non vanishing.

We interpret and address the terms in $H_\mathrm{3}$ as
deformation terms. The remaining contributions, in the case
$R_\theta=0$, give what we call the pantographic Hamiltonian (well
defined in the beginning of next section). In this situation, the
term $H_\mathrm{1}$ can be interpreted as the Hamiltonian of a
particle with a varying mass, while $H_\mathrm{2}$ is the dilation
term. It is anyway important to note that in the presence of
deformation also $H_\mathrm{1} + H_\mathrm{2}$ depends on
$\theta$, hence giving rise to additional deformation terms. In
fact, all the terms in $H_\mathrm{3}$ are associated to
deformation, but there are deformation terms which do not belong
to $H_\mathrm{3}$. This will be clearer in section
\ref{sec:Deformations} where the problem of deformation will be
treated extensively. Instead, in the next section we will consider
only pantographic changes of the domain, hence assuming $R_\theta
= 0$ and $H_\mathrm{3} = 0$.

\section{The pantographic case}\label{sec:Pantography}

We call pantographic Hamiltonian the operator
\begin{equation}\label{eq:PantographicHamiltonian_Def}
  H_\mathrm{p}^\lambda = -\frac{\hbar^2}{2\mu \lambda^2} \nabla^2  +
  \ii\hbar \frac{\dot{\lambda}}{\lambda} \left(1 + \, r \frac{\partial}{\partial
  r}\right)\,,\qquad \partial_\theta\lambda = 0\,,
\end{equation}
i.e. the remaining part of the Hamiltonian in the absence
of deformation, provided $R(t)=\lambda(t)$.

In such a case, the pantographic
Hamiltonian governs the dynamics of the particle in the picture
associated to the fixed domain $\dom_0$. In this simplified
situation it is possible to deduce a rate equation for the energy
of the particle and to derive the exact dynamics in the special
case of constant speed of the walls.

\subsection{The Energy rate equation}

Let us now consider the energy rate equation in the
two-dimensional pantographic case. In the original picture (with
moving walls), the average energy of the system is given by
$E(t)=\Bra{\psi}H\Ket{\psi}$, so that in the picture with static
walls such average energy is given by the mean value, over the
state $\Ket{\phi}$, of the operator $U H U^\dag=H_1+H_3$ in
\eqref{eq:Hamilt_Transformed_Deformation}, from which we have to cancel the
deformation terms, hence obtaining $U H U^\dag = H_\mathrm{1}$.

The energy rate equation is then given by:
\begin{eqnarray}
&&\dot{E}(t) = \frac{\mathrm{d}}{\mathrm{d}t} \Bra{\phi}
H_\mathrm{1} \Ket{\phi} = \frac{i}{\hbar} \Big(
\BraKet{\phi}{H_\mathrm{2}H_\mathrm{1}\phi} -
\BraKet{H_\mathrm{2}H_\mathrm{1}\phi}{\phi}\Big) + \Bra{\phi}
\dot{H}_\mathrm{1} \Ket{\phi}\,,
\end{eqnarray}
with
\begin{eqnarray}
\dot{H}_\mathrm{1} = \frac{\hbar^2 \dot{R}}{\mu R^3} \nabla^2\,.
\end{eqnarray}

It is just the case to mention that the operator
\begin{equation}
H_\mathrm{2}=\ii
\hbar\frac{\dot{R}}{R}\left(1+r\frac{\partial}{\partial
r}\right)=\ii
\hbar\frac{\dot{R}}{R}\left(\frac{1}{2}+r\circ\frac{\partial}{\partial
r}\right),
\end{equation}
where $A\circ B=(AB+BA)/2$ is the Jordan product of the operators
$A$ and $B$, has domain
\begin{equation}
D(H_\mathrm{2})=\{f\in \mathcal{H}^1(\mathbb{R}^2),\;
r\frac{\partial}{\partial r} f(r,\theta)\in L^2(\mathbb{R}^2)\},
\end{equation}
while the product $H_\mathrm{2}H_\mathrm{1}$ has domain:
\begin{equation}
D(H_\mathrm{2}H_\mathrm{1})=\left\{f\in
\mathcal{H}^3(\mathcal{D}_0),\; f(\bold{r})=0\,\, \forall
\bold{r}\in\partial\mathcal{D}_0 \right\}.
\end{equation}
Here we denoted with $\mathcal{H}^1$ the first Sobolev space, i.e.
$\mathcal{H}^1(\mathbb{R}^2)=\{f\in L^2(\mathbb{R}^2) | f'\in L^2(\mathbb{R}^2)\}$,
and with $\mathcal{H}^3$ the third Sobolev space, i.e. the space of square integrable functions with square integrable third derivative.

After introducing $A \equiv r\partial_r$, we can write the rate
equation for the energy in the following way:
\begin{eqnarray}
&&\dot{E} (t) = \frac{\hbar^2 \dot{R}}{2 \mu R^3} \int_{\dom_0} r
\left[ \phi^{*} \left( 1 + A \right) \nabla^2 \phi + \phi \left(
1+A\right)\nabla^2 \phi^{*} + 2 \phi^{*} \nabla^2 \phi \right]
\mathrm{d}r \mathrm{d}\theta \,. \label{eq:Rate_With_A}
\end{eqnarray}

Since the following relations hold, $\left[ A , \vec{\nabla
}\right] = - \vec{\nabla}$,\, $\left[ A , \nabla^2 \right]  = -
2\nabla^2$,\, $\phi^{*} A \nabla^2 \phi = \phi^{*} \nabla^2 A \phi
- 2\phi^{*} \nabla^2  \phi$,\, $\phi A \nabla^2 \phi^{*} = \phi
\nabla^2 A \phi^{*}  -2 \phi \nabla^2  \phi^{*}$,\, the terms
inside the square brackets of \eqref{eq:Rate_With_A} can be
rearranged as $\vec{\nabla} \centerdot \left( \phi^{*}
\vec{\nabla} \phi - \phi \vec{\nabla} \phi^{*} \right)$ $+$
$\phi^{*} \nabla^2 A \phi$ $+$ $\phi \nabla^2 A \phi^{*}$\,.

Now, on the basis of the Gauss-Green theorem and the fact that the
wave function vanishes on the boundary, we can assert that the
divergence does not contribute to the integral in
\eqref{eq:Rate_With_A}. Therefore, we reach the equation
\begin{equation}
\dot{E} (t) = \frac{\hbar^2 \dot{R}}{2 \mu R^3} \int_{\dom_0} r
\left( \phi^{*} \nabla^2 A \phi + \phi \nabla^2 A \phi^{*} \right)
\mathrm{d}r \mathrm{d}\theta \,. \label{eq:Rate_With_A_2}
\end{equation}

By exploiting the Leibniz rule we obtain $\phi^{*} \nabla^2 A \phi
= \vec{\nabla} \centerdot \left( \phi^{*} \vec{\nabla} A \phi
\right) - \left( \vec{\nabla} \phi^{*}\right) \centerdot \left(
\vec{\nabla} (A \phi) \right)$ and  $\phi \nabla^2 A \phi^{*}  =
\vec{\nabla} \centerdot \left( \phi\vec{\nabla} A \phi^{*} \right)
- \left( \vec{\nabla} \phi \right) \centerdot \left( \vec{\nabla}
(A \phi^{*}) \right)$, which, using again the Gauss-Green theorem,
allow to further rearrange the energy rate equation as:
\begin{equation}
\dot{E} (t) = \frac{\hbar^2 \dot{R}}{2 \mu R^3} \int_{\dom_0} r
\left( - \left( \vec{\nabla} \phi^{*}\right) \centerdot \left(
\vec{\nabla} (A \phi) \right) - \left( \vec{\nabla} \phi \right)
\centerdot \left( \vec{\nabla} (A \phi^{*}) \right) \right)
\mathrm{d}r \mathrm{d}\theta \,. \label{eq:Rate_With_A_3}
\end{equation}

Since $\vec{\nabla} \left( A \phi \right) = A \vec{\nabla}  \phi +
\left[ \vec{\nabla},A\right] \phi = A \vec{\nabla}  \phi +
\vec{\nabla} \phi\ $, we get
\begin{equation}
\dot{E} (t) = \frac{\hbar^2 \dot{R}}{2 \mu R^3} \int_{\dom_0} r
\left[ - A \Big( || \vec{\nabla} \phi ||^2\Big) - 2 ||
\vec{\nabla} \phi ||^2 \right] \mathrm{d}r \mathrm{d}\theta \,.
\label{eq:Rate_With_A_4}
\end{equation}

Moreover, integration by part of the first term in the square
brackets gives
\begin{equation}
\int_{\dom_0} r A || \vec{\nabla}\phi||^2 \mathrm{d}r \mathrm{d}\theta=%
\int_{\theta} \int_{r} r^2 \partial_r || \vec{\nabla}\phi||^2 \mathrm{d}r \mathrm{d}\theta %
= \int_{0}^{2 \pi} d \theta \left[ r^2 || \vec{\nabla}
\phi||^2\right]^{r=\eta(\theta)}_{r=0} - 2 \int_{\dom_0} r ||
\vec{\nabla} \phi||^2 \mathrm{d}r \mathrm{d}\theta\,,
\end{equation}
and we eventually get
\begin{equation}
\dot{E}(t) = - \frac{\hbar^2 \dot{R}}{2 \mu R^3} \int_0^{2 \pi}
\mathrm{d}\theta\,  ( r^2 ||\vec{\nabla} \phi||^2 )
\vert_{r=\eta(\theta)} \,.
\end{equation}

It is easy to see that this is the natural $2D$ extension of the
contact term that appears in the one-dimensional case, according
to the analysis developed in Ref~\cite{ref:DiMartino}.

\subsection{The case of a uniformly moving domain}

Let us consider now the special case of a pantographic change with
walls moving at constant velocity, which means $R = 1 + \kappa t$.
In this situation we are able to completely solve the dynamics of
the system by generalizing the results in Ref~\cite{ref:Doescher}.

Assume that a solution of the time-dependent Schr\"odinger
equation is given by a state with the following form:
\begin{eqnarray}
  \phi_n(r, \theta) = e^{\ii P_n(r, t)} \chi_n(r, \theta)\,,
\end{eqnarray}
where $P = \alpha_n(t) r^2 + \beta_n(t)$ is a polynomial of $r$
with time-dependent coefficients and $\chi_n$ is a solution of the
eigenvalue problem $-\hbar^2/(2\mu) \nabla^2 \chi_n = E_n\chi_n$,
with $\phi_n, \chi_n \in L^2[\dom_0]$, $\phi_n(\eta(\theta),
\theta) = \chi_n(\eta(\theta), \theta) = 0$.

We will show that for a suitable choice of $\alpha_n$ and
$\beta_n$ one finds a solution of the Schr\"odinger equation. In
fact, since
\begin{eqnarray}
  \nonumber
  &&\nabla P_n = \frac{\partial P_n}{\partial r} \hat{r} = 2\alpha_n r \hat{r} \,,\\ %
  \nonumber
%  &&(\nabla P_n)^2 = \left(\frac{\partial P_n}{\partial r} \right)^2 = 4\alpha_n^2
%  r^2\,, \\ %
  \nonumber
  &&r \frac{\partial P_n}{\partial r} = 2\alpha_n r^2 \,, \\ %
  \nonumber
  &&\nabla^2 P_n = \frac{1}{r} \frac{\partial P_n}{\partial r} + \frac{\partial^2 P_n}{\partial r^2} = 4\alpha_n\,,%
\end{eqnarray}
the Schr\"odinger equation,
\begin{eqnarray}
  \ii\hbar\partial_t \phi + H_{\mathrm{eff}}\phi = 0\,,
\end{eqnarray}
expressible as,
\begin{eqnarray}
  \nonumber
  \left\{
  - \hbar \dot{P_n} \chi + \frac{\hbar^2}{2 \mu R^2} \left[ \left( \ii (\nabla^2 P_n) - (\nabla P_n)^2 \right)\chi_n + 2 \ii \frac{\partial P_n}{\partial r} \frac{\partial \chi_n}{\partial r} +
  \nabla^2\chi_n \right]
  - \ii \hbar \frac{\dot{R}}{R} \chi_n + \hbar
  \frac{\dot{R}}{R}r \frac{\partial P_n}{\partial r} \chi_n -
  \ii \hbar \frac{\dot{R}}{R}r \frac{\partial \chi_n}{\partial
  r} \right\} e^{\ii P_n} = 0\,,\\
\end{eqnarray}
leads to,
\begin{eqnarray}
  \nonumber
  \left[
  - \hbar \dot{\alpha_n} r^2 - \hbar\dot{\beta_n} + \frac{\hbar^2}{2 \mu R^2} \left( 4 \ii \alpha_n - 4\alpha_n^2 r^2
  \right) - \frac{E_n}{R^2}
  - \ii \hbar \frac{\dot{R}}{R} + 2 \hbar
  \frac{\dot{R}}{R} \alpha_n r^2 \right] \chi_n e^{\ii P_n}
  + \left[ \frac{\ii \hbar^2}{\mu R^2} \frac{\partial P_n}{\partial r}
  - \ii \hbar \frac{\dot{R}}{R} r \right] \frac{\partial
  \chi_n}{\partial r}  e^{\ii P_n}
  = 0\,.\\
\end{eqnarray}
By imposing the coefficient of $\partial \chi_n / \partial r$
equal to zero, one gets
\begin{eqnarray}
  \alpha_n = \frac{\mu}{2\hbar} R \dot{R} \equiv \alpha\,.
\end{eqnarray}
By substituting this result in the coefficient of $\chi_n$, and
imposing it to be zero, one gets:
\begin{equation}
  \left( -\hbar\dot{\beta_n} - \frac{E_n}{R^2} \right) - \frac{\mu}{2} R \ddot{R} r^2 = 0
\end{equation}
which, in the case $\ddot{R} = 0$, gives rise to the solution:
\begin{equation}
  \beta_n = \beta_n(0) - \int_0^t \frac{E_n}{\hbar
  R^2(s)}\mathrm{d}s\,.
\end{equation}
The solution of the Schr\"odinger equation in the original domain
$\dom$ is given by
\begin{eqnarray}
  \psi_n(r, \theta, t) = R^{-1} \, \exp\left(\ii \beta_n(t) + \ii \alpha(t) \left(r / R\right)^2 \right) \, \chi_n(r / R,
  \theta)\,.
\end{eqnarray}
Note that $\beta_n$ depends on the eigenvalue $E_n$ associated to
the state $\chi_n$, while $\alpha_n$ is the same for all
$\chi_n$'s.

Therefore one has
\begin{eqnarray}
  \nonumber
  \int_{\dom} \psi_k^*(r, \theta, t) \psi_l(r, \theta, t) r \mathrm{d}r
  \mathrm{d}\theta &=&
  \int_{\dom} R^{-2} \, e^{\ii (\beta_l(t) - \beta_k(t))} \, \chi_k^*(r / R, \theta) \chi_l(r / R, \theta) \, r \,
  \mathrm{d}r \, \mathrm{d}\theta \\
  &=&
  e^{\ii (\beta_l(t) - \beta_k(t))} \int_{\dom_0} \chi_k^*(s, \theta) \chi_l(s, \theta) \, s \, \mathrm{d}s \, \mathrm{d}\theta = \delta_{kl}\,  \,.
\end{eqnarray}

Then, the $\psi_k$'s are an orthonormal set. Moreover, they
constitute a basis. To prove this statement, consider the
following. It is immediate to prove that since $\chi_k(r,
\theta)$'s are a basis for $L^2[\dom_0]$ then $R^{-1}\chi_k(r/R,
\theta)$'s are a basis for $L^2[\dom]$
--- in both cases with the appropriate boundary conditions. Now,
given any function in $\xi(r, \theta) \in L^2[\dom]$ satisfying
the boundary condition $\xi(R \eta(\theta), \theta) = 0$, one can
consider the function $e^{-\ii \alpha(t) (r/R)^2} \xi(r, \theta)$
and expand it, at time $t$, in terms of the
$R^{-1}\chi_k(r/R, \theta )$'s as follows:
\begin{eqnarray}
  e^{-\ii \alpha(t) (r/R)^2} \xi(r, \theta) = \sum_k c_k \, \chi_k(r/R, \theta) = \sum_k \tilde{c}_k \, e^{\ii
  \beta_k(t)} \, R^{-1} \, \chi_k(r/R, \theta)\,,
\end{eqnarray}
and then,
\begin{eqnarray}
  \xi(r, \theta) = \sum_k \tilde{c}_k \, e^{\ii \alpha(t) (r/R)^2} \, e^{\ii
  \beta_k(t)} \, R^{-1} \, \chi_k(r/R, \theta) = \sum_k \tilde{c}_k \, \psi_k(r, \theta,
  t)\,,
\end{eqnarray}
which completes the proof that the $\psi_k$'s are a basis.

\section{Perturbative treatment of deformations}\label{sec:Deformations}

In this section we try to analyze the case in which deformations
are present. We will concentrate to small deformations, since they
can be treated with a standard time-dependent perturbation
approach.

\subsection{General Framework}

Assume $R(\theta,t) = \lambda(t) [1 + \epsilon f(\theta, t)]$,
where $\epsilon$ plays the role of a deformation parameter
($\epsilon = 0$ means that no deformation is present). In view of
a perturbative treatment, we first assume that $f(\theta, t)$ is a
smooth function and that $\epsilon \ll 1$ (in the end of this
section we comment on such assumptions). Then we consider series
expansions of all relevant functions truncated to the first order
in $\epsilon$: $R^{-2}\approx \lambda^{-2}(1-2\epsilon f)$,
$\dot{R}/R \approx \lambda^{-1}\dot{\lambda} + \epsilon\dot{f}$,
$R^{-1} \approx \lambda^{-1} (1-\epsilon f)$, \, $\partial_\theta
R^{-1} \approx -\epsilon\lambda^{-1}
\partial_\theta f$,\, $\partial^2_\theta R^{-1} \approx -\epsilon\lambda^{-1}
\partial^2_\theta f$. Consequently, we evaluate the first order terms of the total Hamiltonian $H_{\mathrm{eff}}$:
%\begin{subequations}
\begin{equation}
  H_\mathrm{1} = -\frac{\hbar^2}{2 \mu \lambda^2} \nabla^2 + \frac{\hbar^2\epsilon f}{\mu\lambda^2} \nabla^2 + O(\epsilon^2)\,,
\end{equation}
\begin{equation}
  H_\mathrm{2} = \ii\hbar \frac{\dot{\lambda}}{\lambda} \left(1 + \, r \frac{\partial}{\partial
  r}\right) + \ii\hbar \epsilon \dot{f} \left(1 + \, r \frac{\partial}{\partial
  r}\right) + O(\epsilon^2)\,,
\end{equation}
\begin{eqnarray}
  \nonumber
  H_\mathrm{3} &=&
  -\epsilon\,\lambda^{-2}\left(-\frac{\hbar^2}{2\mu}\right)\left[(\partial_{\theta}^2 f) \left( \frac{1}{r^2} + %
  \frac{1}{r} \frac{\partial}{\partial r} \right) + %
  2 (\partial_{\theta} f) \left( \frac{1}{r^2} \frac{\partial}{\partial \theta} + %
  \frac{1}{r} \frac{\partial}{\partial r} \frac{\partial}{\partial \theta} \right)\right] +
  O(\epsilon^2) \\
  &=&
  \epsilon\,\frac{\hbar^2}{2\mu}\,\lambda^{-2}\left[\partial_{\theta}^2 f + 2 (\partial_{\theta} f) \, \frac{\partial}{\partial \theta} \right] \left( \frac{1}{r^2} + %
  \frac{1}{r} \frac{\partial}{\partial r} \right)  +  O(\epsilon^2)%
  \,,
\end{eqnarray}
%\end{subequations}
so that the zeroth order Hamiltonian is $H_\mathrm{eff}^{(0)} =
H_\mathrm{p}^\lambda$, as defined in
\eqref{eq:PantographicHamiltonian_Def}, while the first order
terms are gathered as
\begin{eqnarray}
  H_\mathrm{eff}^{(1)} &=& \epsilon \Bigg\{ \frac{\hbar^2}{\mu}
  \,\lambda^{-2} f\,
  \nabla^2 + \ii\hbar \dot{f} \left(1 + \, r \frac{\partial}{\partial
  r}\right)
  +\frac{\hbar^2}{2\mu}\,\lambda^{-2}\left[\partial_{\theta}^2 f + 2(\partial_{\theta} f) \, \frac{\partial}{\partial \theta} \right] \left( \frac{1}{r^2} + %
  \frac{1}{r} \frac{\partial}{\partial r} \right) \Bigg\}\,.
\end{eqnarray}

Since $\epsilon \ll 1$, we can start by considering the
pantographic dynamics with dilation function $\lambda(t)$ as the
unperturbed dynamics, and then slightly correct the solutions of
the pantographic problem to obtain a first order approximation of
the solutions of the problem with small deformation.
In fact, by applying the standard time-dependent perturbation
theory approach \cite{ref:Sakurai}, after introducing the unitary
operator $U_\mathrm{p}$ such that $\ii\partial_t U_\mathrm{p} =
H_\mathrm{p}^\lambda U_\mathrm{p}$, we get:
\begin{equation}
  \Ket{\psi(t)} = U^\dag \, U_\mathrm{p}(t) \left[ 1 - \ii\hbar\int_0^t U_\mathrm{p}^\dag(s) H_\mathrm{eff}^{(1)}(s) U_\mathrm{p}(s) \mathrm{d}s \right]
  \Ket{\psi(0)}  + O(\epsilon^2) \,,
\end{equation}
where $U$ is the previously introduced unitary transformation that
maps the moving domain into the static one and where we have
exploited the fact that $\Ket{\phi(0)}=\Ket{\psi(0)}$. Moreover,
by introducing unity operators we get the following:
\begin{eqnarray}
  \nonumber
  \Ket{\psi(t)} &=& \sum_\sigma a_\sigma(0) \Ket{\psi_\sigma(t)}
  % \\
  % \nonumber
  % &-&
  - \ii \hbar \sum_{\sigma} \Ket{\psi_\sigma(t)} \int_0^t \Bra{\psi_\sigma(0)} U_\mathrm{p}^\dag(s) U U^\dag H_\mathrm{eff}^{(1)}(s) U U^\dag U_\mathrm{p}(s)
  \sum_{\sigma'} a_{\sigma'}(0) \Ket{\psi_{\sigma'}(0)} \mathrm{d}s
  + O(\epsilon^2) \,\\
  \nonumber
  &=& \sum_\sigma a_\sigma(0) \Ket{\psi_\sigma(t)}
  % \\
  % \nonumber
  % &-&
  - \ii \hbar \sum_{\sigma} \Bigg\{ \sum_{\sigma'} a_{\sigma'}(0) \int_0^t \Bra{\psi_\sigma(s)} U^\dag
  H_\mathrm{eff}^{(1)}(s) U
  \Ket{\psi_{\sigma'}(s)} \mathrm{d}s \Bigg\} \Ket{\psi_\sigma(t)}
  + O(\epsilon^2) \\
  \nonumber
  &=& \sum_\sigma a_\sigma(0) \Ket{\psi_\sigma(t)}
  - \ii \hbar \sum_{\sigma} \Bigg\{  \sum_{\sigma'} a_{\sigma'}(0) \int_0^t \Bra{\phi_\sigma(s)} H_\mathrm{eff}^{(1)}(s)
  \Ket{\phi_{\sigma'}(s)} \mathrm{d}s \Bigg\} \Ket{\psi_\sigma(t)}
  + O(\epsilon^2)\,, \\
\end{eqnarray}
where $\sigma$ and $\sigma'$ denote suitable sets of quantum
numbers, and $a_\sigma(0) = \BraKet{\psi_\sigma(0)}{\psi(0)}$.

It is necessary to stress here the importance of evaluating the
action of $U$ exactly, in order to prevent violation of the
boundary conditions.

Moreover, at this stage, it is worthy to spend some words about the
validity of our perturbation treatment itself, since one could wonder
whether all the neglected terms of the Hamiltonian and the discarded
terms in the perturbative expansion are really negligible.
A positive answer is based on the treatment of {\it
boundary perturbation} shortly reported in \cite{ref:Kato}, where
it is introduced the problem of a \lq\lq free\rq\rq\, particle in
a box whose shape is slightly modified. According to Kato,
provided the deformation is \lq\lq small\rq\rq\, and \lq\lq
smooth\rq\rq\,, Taylor expansion of the eigenvalues and
eigenvectors is justified and perturbation treatment is
legitimated. In our case, we have two additional ingredients: (a)
a dilation is also considered, not simply a deformation, and (b)
the perturbation is time dependent; but none of such two elements
introduce significant differences. In fact, the presence of the
dilation implies that the condition of \lq\lq smallness\rq\rq\, is
strictly fulfilled provided $\epsilon\lambda(t) \sup_{\theta}
f(\theta, t) \ll 1$ $\forall t$, while the \lq\lq
smoothness\rq\rq\, is guaranteed by the smoothness of the
deformation function $f(\theta, t)$ with respect to $\theta$.
Finally, the time dependence implies the presence of a transport
term ($\ii\hbar\dot{U} U^\dag$), whose structure, in the specific
case, gives rise to the dilation term ($\propto
\alpha(\theta)(1+r\partial_r)$), and again, following the line of
\cite{ref:Kato} it turns out that, under the assumption of smooth
and small deformation, one easily shows that this additional term
does not compromise the possibility of Taylor-expanding the
eigensolutions and to resort to perturbation treatment.

\subsection{The case of an elliptical box}

As an example, let us consider a circular box which expands and
squeezes becoming an ellipse.

The moving and static domain are:
\begin{equation}
  \dom = \left\{ (r, \theta): r \le {r_0 \lambda(t)}/{(1-\epsilon g(t)\cos\theta)}\,, \theta \in [0, 2\pi] \right\} \,,\qquad
  \epsilon \ll 1\,,
\end{equation}
\begin{equation}
  \dom_0 = \left\{ (r, \theta): r \le r_0\,, \theta \in [0, 2\pi]  \right\}\,,
\end{equation}
where $\lambda(t)$ and $g(t)$ are suitable functions such that
$\lambda(0)=1$ and $g(0)=0$ (at $t=0$ the domain $\dom$ coincides
with $\dom_0$), and $g(t)$ tends toward unity (after the value of
$g(t)$ stabilizes the shape of $\dom$ does not significantly
change). Here $\epsilon$ is the asymptotic value of the
eccentricity of the time-changing ellipse, and since it is much
smaller than unity then we get:
\begin{equation}
  R(\theta, t)= \frac{\lambda(t)}{1-\epsilon f(\theta, t)} \approx \lambda(t)
  [1+\epsilon f(\theta, t)]\,,\qquad f(\theta, t) =
  g(t) \cos\theta\,,
\end{equation}
\begin{eqnarray}\label{Ham}
  % \nonumber
  H_\mathrm{eff}^{(1)} = %&=&
  \epsilon \left[
  \frac{\hbar^2}{\mu} g \lambda^{-2} \cos\theta\,
  \nabla^2 + \ii\hbar \dot{g} \cos\theta \left(1 + \, r \frac{\partial}{\partial
  r}\right) % \\
  - \, \frac{\hbar^2}{2\mu}\,g\, \lambda^{-2} \,
  \left(\cos\theta + 2 \sin\theta\frac{\partial}{\partial\theta}\right) \left(\frac{1}{r^2} + %
  \frac{1}{r} \frac{\partial}{\partial r} \right)
  \right]%
  \,.
\end{eqnarray}

Let us consider the case in which the gross evolution (the
pantographic one) is associated to uniformly dilating box --- this
means $\lambda(t) = 1+\kappa t$ --- since in this case we know the
solution of the relevant pantographic problem. Moreover, let us
assume a smooth form for the function $g$, for example
$g(t)=1-e^{-\gamma t}$.

The energy eigenvalues and stationary states of a quantum particle
confined in a circle of radius $r_0$ are given
by~\cite{ref:Robinett1996}:
%\begin{subequations}
\begin{eqnarray}
  E_{mn} &=& \frac{\hbar^2}{2 \mu r_0^2} a_{mn}^2\,,\\
  \chi_{mn} &=& (2\pi)^{-1/2} A_{mn} J_{m}(k_{mn} r ) \times e^{\ii m \theta}\,, \qquad m \in
  \mathbb{Z}\,,
\end{eqnarray}
where $J_m(x)$ is the Bessel function of $m$-th order, $a_{mn}$ is
the $n$-th zero of $J_m$, and
\begin{eqnarray}
  && k_{mn}^2 = \frac{2 \mu E_{mn}}{\hbar^2} = \frac{(a_{mn})^2}{r_0^2} \; \Rightarrow \; k_{mn} = \frac{|a_{mn}|}{r_0} \, ,\\
  && A_{mn} = \left(\int_0^{r_0} r J_m (k_{mn}r)^2
  \mathrm{d}r\,\right)^{-1/2}\,.
\end{eqnarray}
%\end{subequations}

It is very useful to note that in this case we have, for the
matrix element $\Bra{\phi_{mn}(s)} H_\mathrm{eff}^{(1)}(s)
\Ket{\phi_{m'n'}(s)}$, the selection rule $m = m'\pm 1$. In the
appendix \ref{app:MatrixEl} we give the explicit expressions of
the matrix elements involved in the perturbative treatment.

The selection rule is useful both as a simplification in the
evaluation of the dynamics and as a warranty of the stability of
the perturbation treatment, in this case. In fact, among the terms
in Eq.(\ref{Ham}) there are unbounded operators involving the
variable $r$ (they are, $r^{-2}$ and $r^{-1}\partial_r$), but due
to the behavior of the Bessel functions close to zero ($J_m(r)\sim
r^m$ when $r\rightarrow 0$), the only divergent matrix elements of
the operators involving the radial variable are those between two
$J_0$'s: $\int_0^1 r J_0(k_{0n}r) J_0(k_{0n'}r) \mathrm{d}r$. Now,
since such Bessel functions correspond to the same angular
momentum ($m=m'=0$), the condition $m=m'\pm 1$ is not fulfilled
and the relevant matrix element of $H_\mathrm{eff}^{(1)}$ is zero.
Therefore $H_\mathrm{eff}^{(1)}$ is \lq effectively bounded\rq\,,
in spite of the presence of unbounded operators related to the
radial part.

\begin{figure}[t!]
\centering
\includegraphics[width=0.35\textwidth, angle=0]{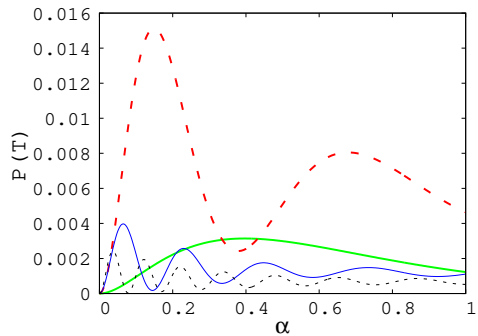} %
\caption{(Color online). Populations of the states
$\Ket{\psi_{1,1}}$ (green bold solid line), $\Ket{\psi_{1,2}}$ (red bold dashed line),
$\Ket{\psi_{1,3}}$ (blue solid line), $\Ket{\psi_{1,4}}$ (black dashed line),
when the initial
state is $\Ket{\psi_{0,1}}$. Here the time is expressed in
$\kappa^{-1}$ units, $\hbar=1$, $\epsilon=0.05$ and
$\gamma=5\kappa$.} \label{fig:transition_probab_1}
\end{figure}

\begin{figure}[t!]
\centering
\includegraphics[width=0.35\textwidth, angle=0]{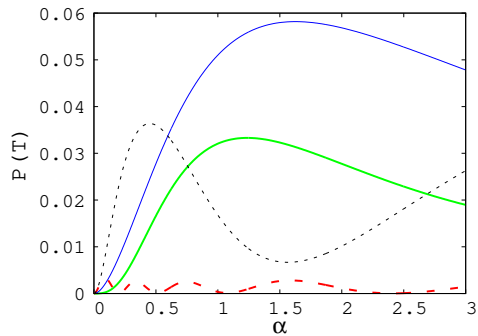} %
\caption{(Color online). Populations of the states
$\Ket{\psi_{3,1}}$ (green bold solid line), $\Ket{\psi_{3,2}}$ (red
bold dashed line), $\Ket{\psi_{1,1}}$ (blue solid line),
$\Ket{\psi_{1,2}}$ (black dashed line), when the initial
state is $\Ket{\psi_{2,1}}$. Here the time is expressed in
$\kappa^{-1}$ units, $\hbar=1$, $\epsilon=0.05$, and
$\gamma=5\kappa$.} \label{fig:transition_probab_2}
\end{figure}

In the following, we show some numerical calculations of the
transition probabilities associated to specific initial
conditions.
We first consider the initial state
$\Ket{\psi(t=0)}=\Ket{\psi_{0,1}}$ (the ground state), so that, at
first order in the eccentricity, only transitions to states
$\Ket{\psi_{\pm 1, n}}$ are allowed. In particular, in
Fig.\ref{fig:transition_probab_1} we show the transition
probabilities to the states $\Ket{\psi_{1,n}}$ with $n=1,2,3,4$.
Due to the specific structure of the initial state and of the
states of the basis, the transition probabilities to
$\Ket{\psi_{1, n}}$ and $\Ket{\psi_{-1, n}}$ are equal.
Moreover, in Fig.\ref{fig:transition_probab_2} we show some of the
transition probabilities evaluated assuming $\Ket{\psi_{2, 1}}$ as
the initial condition.

\section{Discussion}\label{sec:Discussion}

In this paper we have extended the technique of
Ref~\cite{ref:DiMartino} to the cases of two-dimensional boxes. We
have first of all reviewed the results related to the
one-dimensional problem, and then we have studied in depth
the case of a \lq\lq free\rq\rq particle confined in a
two-dimensional box by mapping such problem into the one of a
particle in a fixed domain but governed by a time-dependent Hamiltonian.

The two-dimensional problem reveals an important element of
novelty with respect to the one-dimensional case, which is the
role of the shape of the box (and its changing) through the whole
process. In fact, the time-dependent Hamiltonian governing the
dynamics of the particle in the picture associated to a fixed
boundary is the sum of several terms, many of which disappear when
the boundary of the box changes its dimensions but not its shape
(we have addressed this situation as the pantographic case). This
has allowed to single out two classes of 2D problems depending on
the presence of deformations of the contour of the box.

On the one hand we have shown that in the absence of deformation
the results obtained in the one-dimensional case can be naturally
extended. For example, the rate equation for the energy is just
what one could guess starting from the rate equation of the energy
in the one-dimensional moving box. Moreover, the time evolution of
the particle in the case of uniformly moving boundaries is the
very natural extension of the one-dimensional counterpart.

On the other hand when we consider deformation of the boundary,
the situation becomes more complicated and the previously reported
analogies to the one-dimensional case are no longer valid. In
fact, the additional terms of the Hamiltonian coming from the
deformation of the contour make the resolution of the relevant
dynamical problem very difficult. We have then used an approach
based on perturbation theory assuming the pantographic Hamiltonian
as the unperturbed one, and the deformation terms, which play the
role of perturbation, are assumed to be small enough. On this
ground, we have shown that deformation of the boundary is
responsible for transitions between pantographic states, which
would not occur otherwise.

In order to conclude our analysis, we want to briefly comment on
the fact that the methods and results reported for the 2D case can
be easily extended to the case of a particle in a
three-dimensional box. Indeed, in the 3D pantographic case, after
performing the passage to the static boundary picture, the
generator of time evolution turns out to be again the sum of
the kinetic energy of a varying-mass particle and a dilation term
very similar to the two-dimensional one. The complete dynamical 
solution that one can obtain when the velocity of the boundary is 
constant and the general energy rate equation (valid for any time 
dependence of the velocity of the walls) are the very
natural extensions of the two-dimensional counterparts. Finally,
if one considers also the presence of deformation, this can be
treated perturbatively by following the same approach previously
developed. Obviously, every calculation is more cumbersome, since
it involves more terms than in the two-dimensional case.

\section{Acknowledgements}

We thank P. Facchi, G. Marmo and S. Pascazio for stimulating
discussions on the subject of this paper. S.D.M. acknowledge support from the Italian National Group of Mathematical Physics (GNFM-INdAM).

\newpage

\appendix

\section{Matrix Elements for the perturbative treatment}\label{app:MatrixEl}

In this appendix we give the explicit result of the evaluation of
the matrix elements of $\hat{H}_{\mathrm{eff}}^{(1)}$ for a circular box which is deformed to an elliptical box and expanded.

Since
\begin{eqnarray}
&&\hat{H}^{(1)}_{\mathrm{eff}} = \hat{H}_{1}^{(1)} +
\hat{H}_{2}^{(1)} + \hat{H}_{3}^{(1)}\,,
\end{eqnarray}
with obvious notation, we give separately the matrix elements of
the three operators:

\begin{eqnarray}
\nonumber
&& \int_0^t \Bra{\phi_{mn}}H_\mathrm{1}^{(1)}(s)\Ket{\phi_{m'n'}} \mathrm{d}s  = \epsilon \left( \delta_{m,\mmp + 1} + \delta_{m,\mmp - 1} \right)  \times \Large[ 2 \FTW \times \WTW \\
&& + \, \FTH \times \WTH + \Bigg. 2  \FTW \times \WF -
(k_{m'n'})^2 \FO \times \WTW \Large]\,,
\end{eqnarray}

\begin{eqnarray}
\nonumber
&& \int_0^t \Bra{\phi_{mn}}H_\mathrm{2}^{(1)}(s)\Ket{\phi_{m'n'}} \mathrm{d}s  =   \epsilon  \left( \delta_{m,\mmp + 1} + \delta_{m,\mmp - 1} \right)  \times \\
&& \left[ \FF  \times \WO +  \FFI \times \WTH  + \FF  \times \WF
\right]\,,
\end{eqnarray}

\begin{eqnarray}
&& \int_0^t \Bra{\phi_{mn}}H_\mathrm{3}^{(1)}(s)\Ket{\phi_{m'n'}} \mathrm{d}s  =  \epsilon \left[ \left( 1/2 + m' \right)  \delta_{m,\mmp + 1} + \left( 1/2 - m' \right)\delta_{m,\mmp - 1} \right] \times \\
&& \left[\FO \times W^{(1)}_{mn,m'n'} + F^{(2)}_{mn,m'n'} (t)
\times W^{(2)}_{mn,m'n'} \nonumber \right]\,,
\end{eqnarray}

where:

\begin{subequations}
\begin{eqnarray}
&&\FO \equiv \frac{\hbar^2}{2 \mu}  \int_0^t \frac{g(s)}{\lambda^2 (s)} \mathrm{Exp} \left[ i\xi^{mn}_{\mmp \nnp}(s)\right] \, \mathrm{d}s\,,\\
&&\FTW = % \frac{\hbar^2}{2\mu} \int_0^t \frac{g(s)}{\lambda^2 (s)} \mathrm{Exp} \left[ i\varphi^{mn}_{\mmp \nnp}(s)\right]  2i \alpha (s) \, \mathrm{d}s  =
i \hbar \int_0^t g(s) \frac{\dot{\lambda} (s)}{\lambda (s)} \mathrm{Exp} \left[ i\xi^{mn}_{\mmp \nnp}(s)\right]\,\mathrm{d}s\,,\\
&&\FTH = % \frac{\hbar^2}{2\mu} \int_0^t \frac{g(s)}{\lambda^2 (s)} \mathrm{Exp} \left[ i\varphi^{mn}_{\mmp \nnp}(s)\right]  (2i \alpha (s))^2 \, \mathrm{d}s =
- \mu \int_0^t g(s) \dot{\lambda}^2(s) \mathrm{Exp} \left[ i\xi^{mn}_{\mmp \nnp}(s)\right]\,\mathrm{d}s\,,\\
&&\FF  \equiv \frac{\ii \hbar}{2} \int_0^t  \dot{g}(s) \mathrm{Exp} \left[ i\xi^{mn}_{\mmp \nnp}(s)\right]\,\mathrm{d}s\,,\\
&&\FFI = % \frac{\ii \hbar}{2} \int_0^t \dot{g}(s) \mathrm{Exp} \left[ i\varphi^{mn}_{\mmp \nnp}(s)\right]   2i\alpha (s) \, \mathrm{d}s =
- \frac{\mu}{2} \int_0^t \dot{g}(s) \lambda(s) \dot{\lambda} (s)
\mathrm{Exp} \left[ i\xi^{mn}_{\mmp \nnp}(s)\right]\,\mathrm{d}s
\,,
\end{eqnarray}
\end{subequations}
with
\begin{eqnarray}
\xi^{mn}_{\mmp \nnp}(t) = \beta_{m'n'}(t)-\beta_{mn}(t) \,,
\end{eqnarray}
and
\begin{subequations}
\begin{eqnarray}
&&\WO = A_{mn} A_{m'n'}\,  \int_0^{r_0}   \Bess \left(\frac{1}{r} + \frac{\partial}{\partial r}\right) \Bessp \, \mathrm{d}r \,. \\
&&\WTW = A_{mn} A_{m'n'}\, \int_0^{r_0}r \Bess \Bessp \, \mathrm{d}r \,,\\
&&\WTH = A_{mn} A_{m'n'} \int_0^{r_0}  r^3 \Bess  \Bessp \, \mathrm{d}r \,,\\
&&\WF = A_{mn} A_{m'n'} \int_0^{r_0}  r^2 \Bess
\frac{\partial}{\partial r} \Bessp \, \mathrm{d}r \,.
\end{eqnarray}
\end{subequations}

\end{document}